\documentclass[aps,12pt]{revtex4}
\usepackage{epsfig}
\usepackage{graphicx}
\usepackage{subfigure}
\usepackage{ulem}
\usepackage{amsfonts}
\usepackage{epsfig,bm}
\usepackage{comment}
\newcommand{\be}{\begin{equation}}
\newcommand{\ee}{\end{equation}}
\newcommand{\ba}{\begin{eqnarray}}
\newcommand{\ea}{\end{eqnarray}}

\begin{document}
\title{On Attractor Flow and Small Black Holes}
\vspace{0.50cm}
\author{Sanjay Siwach$^{a}$}
\email{sksiwach@hotmail.com}
\author{Bhupendra Nath Tiwari$^{b}$} \email{bntiwari.iitk@gmail.com} 
 
\affiliation{$^{a}$ Department of Physics,\\
Banaras Hindu University,\\
Varanasi- 221 005, India.\\}

\affiliation{$^{b}$INFN-Laboratori Nazionali di Frascati\\
Via E. Fermi, 40 -- I-00044 Frascati\\
Rome, Italy.} 

\vspace{1.0cm}
\noindent
\begin{abstract}
We study the attractor flow and near horizon geometry of
two-charge small black holes in heterotic string theory. The
Hessian of Sen's entropy function with respect to the moduli
fields has standard attractor properties and shows the
interesting factorization at the attractor fixed points.
We notice that the stability conditions are preserved under
arbitrary $\alpha^{\prime}$-corrections to the black hole
solutions.
\end{abstract}
 
\maketitle
\setcounter{page}{0}

\newpage

Black holes in string theory are the solutions of low energy
supergravity actions. They obey the laws of
black hole thermodynamics just like the black holes solutions in
general theory of relativity. In particular, the macroscopic
entropy is proportional to horizon area. It has been possible to count
microstates for certain kind of black holes in string theory and
in the large charge limit, the statistical entropy
\cite{Strominger:1996,msw} coincides with Bekenstein-Hawking entropy given
in terms of the horizon area. The subleading higher derivative
corrections are important particularly for small black holes for
which the leading Bekenstein-Hawking entropy
vanishes \cite{Sen:1995in1, Sen:1995in2}.

Attractor mechanism \cite{att1,att2,att3,att4} plays an important
role in understanding the physics of string theory black holes.
The scalar fields starting from their generic boundary values at
radial infinity coupled to the black hole evolve towards their
fixed values at the horizon. The fixed values are determined by
the minima of black hole effective potential, which in general
depends on the scalar fields and a set of invariant
electric-magnetic charges. The attractor configurations, away from
the critical points of the black hole potential also involve
moduli fields at the asymptotic infinity \cite{9807108}. The
sensitivity of attractor flow to the higher derivative $
\alpha^{\prime}$-corrections leads to the stringy insights into
the moduli behavior away from attractor fixed points \cite{osv,p}.
This involves the effect of all possible field fluctuations
against the macroscopic attractor configuration.

Entropy function formalism
\cite{Sen1,Sen2,Sen3,Sen4,Sen5,Sen6,s,dst} seems to capture all
higher derivative corrections to black hole entropy. The stuty of
generalized attractor properties and associated nature of moduli
in this formalism offers an unified interrelation of (i) black
hole attractors, (ii) stringy $\alpha^{\prime}$-corrections and
(iii) behavior of moduli flow equations. In general, this picture
indicates non-trivial statistical configurations of the charges
and moduli fields. In the present article, we shall focus our attention
on the small black holes, and explicate the nature of attractor
flow as moduli correlations.

We investigate the higher derivative $\alpha^{\prime}$-corrections
for small blck holes from the perspective of Sen's entropy function.
The attractor equations follow directly from the extremization of
Sen's entropy function \cite{Sen1,Sen2,Sen3,Sen4,Sen5,Sen6,s,dst}
for a given set of higher derivative
$\alpha^{\prime}$-corrections. The flow equations arising from the
Hessian of Sen's entropy function of the small black holes have been
analyzed with respect to attractor moduli. The definition of the
Hessian provides information about the covariant attractor flow on
the moduli space.

In present paper, we shall examine these notions for two charge small 
black holes in heterotic string theory with arbitrary
stringy corrections. The attractor properties of moduli are
modified in general, but the attractor fixed point macroscopic
configurations remain the same, in the present case. The higher
order $\alpha{\prime} $-corrections do not enter in the Hessian
function $ F_{ij}(u_S,u_T) $ of corrected entropy functions at any
order for small black holes. We notice that Sen's entropy
function and attractor flow at fixed points to be in 
generic agreement with the nature of the Hessian of BPS-mass
factorization \cite{fgk}.

The extremal black holes in four spacetime dimensions
\cite{fgk,fk,gijt,k,tt0,g,kss,aft} feature a near horizon geometry
of the form $AdS_2\times S^2$. Thus, the area of the horizon can
be thus expressed in terms of (i) radius of $S^2$, (ii) electric
and magnetic charges, which in turn are determined as the minimum
of the black hole effective potential $V_{BH}$, by the attractor
mechanism \cite{fgk,fk,gijt,k,tt0,g,kss,aft}. It is interesting to
analyze the attractor behavior of $V_{BH}(p,q,\phi^a)$, defined as
the function of moduli fields and charges of the theory. To
include the effect of asymptotic moduli fields \cite{fgk}, a
generalization has been proposed as the negative Hessian function
of the effective potential. The positivity of the Hessian provides
stability condition leading to the critical behavior
\cite{fgk,fk,gijt,k,tt0,g,kss,aft}, at the extremum of $
V_{BH}$.

We shall analyze black hole macroscopic configuration away from
the attractor fixed point(s) by defining the flow equations as the
negative Hessian of Sen entropy function;
\begin{eqnarray}
g_{ij}^{(F)}=  -\bigtriangledown_i \bigtriangledown_j\ F(p, q,
u_i, v_i)
\end{eqnarray}
We shall be interested in attractor flow equations of the small
black holes in heterotic string theory including all order
corrections in $\alpha{\prime}$. Considering the near horizon
geometry of the form of $AdS_2\times S^{D-2}$, the typical field
configuration takes the following form;
\begin{eqnarray} \label{horgeom}
ds^2 &=& v_1(-r^2dt^2+ \frac{dr^2}{r^2})+ v_2 d\Omega_{D-2}^2 \nonumber \\
S &=& u_S, ~~~ T = u_T \nonumber \\
F_{rt}^{(i)} &=& e_i; \ i= 1,2
\end{eqnarray}
Let us briefly summarize the Sen's entropy function formalism. For
a generically covariant Lagrangian density $\mathcal L $, we
define the horizon function;
\begin{equation}
f(\overrightarrow{u}, \overrightarrow{v}, \overrightarrow{e})=
\int_{S^{D-2}}\sqrt{-g} \ \mathcal L,
\end{equation}
such that, the equations of motion in near horizon geometry
Eqn.$(\ref{horgeom})$ read as;
\begin{equation}
\frac{\partial f}{\partial u_i}= 0, ~~~ \frac{\partial f}{\partial
v_j}= 0 ;
\end{equation}
The electric charges of the theory are defined as
\begin{equation} \label{echarges}
q_i:= \frac{\partial f}{\partial e_i}
\end{equation}
Then, Sen's entropy function
\cite{Sen1,Sen2,Sen3,Sen4,Sen5,Sen6,Sen7,Sen8,Sen9} takes the
form;
\begin{equation} \label{sef}
F(u_i, v_j, q_k)= 2 \pi (\sum_{l=1,2} e_lq_l- f(u_i, v_j,e_l)).
\end{equation}
Before proceeding further, it is worth to mention that there is
nontrivial and unique choice of the small black hole parameters,
independent of $D$, for which black hole entropy reduces to the
statistical entropy $S= 4\pi\sqrt{nw}$, where $n$ is KK momentum
and $w$ is winding charge. At finite order
$\alpha^{\prime}$-corrections, there exists exact matching of the
macroscopic attractor entropy and microscopic statistical entropy
of the small black holes \cite{Sen8}. The analysis follows from
consideration of 1/2-BPS states of heterotic string
configuration compactified on $T^{9-D}\times S^1$ \cite{d,ddmp}.

For $AdS_2\times S^{D-2}$ near horizon geometry, Prester
\cite{Prester} has shown that the horizon function reduces to
\begin{eqnarray}
f(u_i, v_j,e_k)&=& b u_Sv_1v_2^{(D-2)/2} \lbrace \frac{2
u_T^2e_1^2}{v_1^2}+ \frac{2e_2^2}{u_T^2v_1^2} \nonumber \\ && +
\sum_{m=1}^{[D/2]} ( \alpha^{ \prime m-1} \lambda_m
\frac{(D-2)!}{(D-2m)!}v_2^{-m}  \nonumber \\ && [(D- 2m)
(D-2m-2)-2m\frac{v_2}{v_1}]\rbrace),
\end{eqnarray}
where $\lambda_1= 1 $ and $ b= \frac{\Omega_{D-2}}{16 \pi G_N}$.

Consider the simplest case with $\lbrace e_i \rbrace
\leftrightarrow \lbrace q_i \rbrace$. In $D=4$ and $D=5$ spacetime
dimensions, this implies that the Sen entropy function remains
unchanged for the small black holes with the following horizon
function \cite{Prester}
\begin{eqnarray}
f(u_S,u_T, v_1, v_2,e_1,e_2)&=& b u_Sv_1v_2^{(D-2)/2}
[\frac{2 u_T^2e_1^2}{v_1^2}+ \frac{2e_2^2}{u_T^2v_1^2} \nonumber \\
&-& \frac{2}{v_1} \frac{(D-2)(D-3)}{v_2}(1- \frac{4
\alpha^{\prime} \lambda_2 }{v_1})]
\end{eqnarray}
Using Eqn.$(\ref{echarges})$, we find that the associated
charges of the theory are
\begin{eqnarray}
q_1&=& \frac{\partial f}{\partial e_1}=
4b u_Sv_1^{-1}v_2^{(D-2)/2}u_T^2e_1,  \nonumber \\
q_2&=& \frac{\partial f}{\partial e_2}= 4b
u_Sv_1^{-1}v_2^{(D-2)/2}u_T^{-2}e_2
\end{eqnarray}
For the small black hole in the above spacetime dimensions, it
follows from Eqn.$(\ref{sef})$ that Sen entropy function has the
following form;
\begin{eqnarray} \label{sef45}
F(u_S,u_T, v_1, v_2,q_1,q_2)&=&2 \pi [
\frac{v_1}{8bu_Sv_2^{(D-2)/2}u_T^2} q_1^2+
\frac{v_1u_T^2}{8bu_Sv_2^{(D-2)/2}} q_2^2 + 2b u_Sv_2^{(D-2)/2}
\nonumber \\ &-& b u_Sv_1v_2^{(D-2)/2}\frac{(D-2)(D-3)}{v_2}(1-
\frac{4 \alpha^{\prime} \lambda_2 }{v_1})]
\end{eqnarray}
The corresponding Hessian derivatives of the above entropy
function with respect to the moduli fields are;
\begin{eqnarray}
g_{SS}&=& 2 \pi ( \frac{1}{4} \frac{v_1 q_1^2}{bS^3v_2^{(D/2-1)}T^2}+ \frac{1}{4} \frac{v_1T^2q_2^2}{bS^3v_2^{(D/2-1)}}),\nonumber \\
g_{ST}&=& 2 \pi ( \frac{1}{4} \frac{v_1 q_1^2}{bS^2v_2^{(D/2-1)}T^3}- \frac{1}{4} \frac{v_1Tq_2^2}{bS^2v_2^{(D/2-1)}}),\nonumber \\
g_{TT}&=& 2 \pi ( \frac{3}{4} \frac{v_1
q_1^2}{bSv_2^{(D/2-1)}T^4}+ \frac{1}{4}
\frac{v_1q_2^2}{bSv_2^{(D/2-1)}}),
\end{eqnarray}
where $S$ and $T$ denote the moduli fields $u_S$ and $u_T$. In
sequel, we also denote these flow components as $g_{11}$,
$g_{12}$, and $g_{22}$. In this case, it is further known
\cite{Prester} that the attractor equations lead to the following
horizon values of the spacetime parameters and moduli fields;
\begin{eqnarray} \label{eom45}
v_1 &=& 4 \alpha^{\prime} \lambda_2, \nonumber \\
v_2 &=& 4 (D-2)(D-3) \alpha^{\prime} \lambda_2m, \nonumber \\
u_T &=& \sqrt{\frac{n}{w}} ,\nonumber \\
u_S &=& \frac{4 \pi \alpha^{\prime}
G_N}{\Omega_{D-2}}\frac{v_1}{v_2^{(D-2)/2}}
\sqrt{\frac{2nw}{\lambda_2}}
\end{eqnarray}
The corresponding value of electric fields are given as
\begin{eqnarray}
e_1&=& \sqrt{2 \alpha^{\prime} \lambda_2 \frac{n}{w}} \nonumber \\
e_2&=& \sqrt{2 \alpha^{\prime} \lambda_2 \frac{w}{n}}
\end{eqnarray}
Using the attractor values, viz., Eqn.$(\ref{eom45})$, it has been
known \cite{Prester} that the entropy of small black hole, as
attractor value of Eqn.$(\ref{sef45}$), is given by, $ S = 4\pi
\sqrt{8 \lambda_2} \sqrt{nw}$. This further matches with the
statistical entropy $S = 4\pi \sqrt{nw}$ of corresponding
microscopic configuration of counting the underlying string
states, if $ \lambda_2 = \frac{1}{8}$.

At the above attractor values of moduli fields, we find that the
generalized attractor flow equations arising from Eqn.$(\ref{sef45})$ are
\begin{eqnarray}
g_{11}&=& \frac{1}{32} \frac{\Omega_{D-2}^2 \sqrt{16} (q_1^2 w^2+
n^2 q_2^2)(\frac{1}{4} \alpha^{\prime} D^2-
\frac{5}{4} \alpha^{\prime} D+ \frac{3}{2} \alpha^{\prime})^D} {\pi \alpha^{\prime} (D-2)^2 (D-3)^2 G_N^2 n^2 w^2 \sqrt{nw}}, \nonumber \\
g_{12}&=& -\frac{1}{4} \Omega_{D-2} \frac{(-q_1^2 w^2+ n^2 q_2^2)
(\frac{1}{4} \alpha^{\prime} D^2-
\frac{5}{4} \alpha^{\prime} D+ \frac{3}{2} \alpha^{\prime})^{D/2}}{(D-2) (D-3) G_N n^2 w^2} \sqrt{\frac{w}{n}},\nonumber \\
g_{22}&=& \pi \frac{\alpha^{\prime}}{n^2} \sqrt{\frac{2
\lambda_2}{nw}}(3 q_1^2 w^2+ n^2 q_2^2)
\end{eqnarray}
Therefore, we see that the $TT$ component of the Hessian of the
Sen's entropy function remains unchanged, under the change of dimensionality. 
At this point, it is noteworthy to mention
from the geometric perspective of entropy function \cite{bntESC}
that the small black hole thermodynamics, arising from the
attractor fixed point configurations, show a degenerate system
\cite{bnt,SST,BNTBull1,BNTBull2}.

What follows further that the Hessian of Sen entropy function with
respect to the moduli fields $\lbrace S, T \rbrace$ has standard
attractor horizon properties. When it is evaluated at the an
attractor fixed point of the small black hole configuration, one
obtains the following factorization
\begin{eqnarray} \label{moduli}
\left (\begin{array}{rr}
    g_{SS} & g_{ST} \\
     g_{ST} & g_{TT} \\
\end{array} \right)_{attr}
= K_{IJ}( S_{attr}, T_{attr}) S_{BH}(n,w);
\end{eqnarray}
where $\{K_{IJ}; I, J = S, T \}$ signify the second derivative of
the Sen entropy function, and the subscripts S and T, as before,
are understood as the respective partial derivatives with respect
to the moduli fields $u_S$ and $u_T$. This is the key result of
this paper. Similar analysis arises from the standard attractor
equations in the viewpoint of black hole effective potential
energy. It has been mainly introduced in standard supergravity
literature by Ferrara, Kallosh and Strominger
\cite{att1,att2,att3,att4,fgk}.

We anticipate that the higher derivative corrections do not
contribute into the Hessian of the entropy function$ F(u_S,u_T) $, and thus into
the flow equations. This follows from the fact that the moduli $
\lbrace S, T \rbrace $ appear only linearly in the higher
derivative corrected small black hole entropy functions.
Furthermore, we find that the higher order $ \alpha{\prime}
$-corrections arising from the heterotic string compactification
\cite{d,ddmp} on $ T^{9-D}\times S^1 $ also do not enter in the Hessian
matrix $ F_{ij}(u_S,u_T) $ of entropy functions at any finite
order.

Notice that the moduli aspects of solution which we show here is
true for arbitrary $D$ dimensional small black holes,
\textit{viz.}, the components $ \lbrace K_{IJ}; I,J= S,T \rbrace $
determine the moduli space properties of underlying heterotic
string compactifications. In general, a $D$ dimensional spacetime
passing from the odd dimensions $D= 2m-1$ to the even dimensions
$D=2m$ and then to the odd dimensions $D= 2m+1$, it is known
\cite{Prester} that the horizon function $f$ receives the
following additional contribution
\begin{eqnarray} \label{corr}
\Delta f = \frac{\Omega_{D-2}}{16\pi G_N} u_S v_1 v_2^{(D-2)/2}
\alpha'^{m-2} \frac{(D-2)!}{(D-2m)!} v_2^{-m+1} \left(
\lambda_{m-1} - \frac{2m\alpha'}{v_1} \lambda_m \right),
\end{eqnarray}
where one supposes for all $k=1,\ldots,m-1$ that $\lambda_{k}$ may
be determined from the lower-dimensional analysis. Thus,
$\lambda_m$ is the only free parameter to be determined at the
chosen order of corrections. We notice further that the correction
term Eqn.$(\ref{corr})$ is linear in $u_S$ moduli. We find that
such contributions do not manifestly enter into the Hessian of
arbitrary higher derivative corrected entropy function.

Thus, the generalized flow equations as the equations of motion of
Sen entropy function, and thereby factorization property of the
Hessian are independent of dimensionality of the small black hole
solutions. From the gravity side, we have shown that
$\alpha^{\prime}$-corrections remain intact against the
thermodynamic nature of moduli space attractor configurations.
Such considerations may be envisaged to be of central importance
in statistical physics \cite{BNTBull1, BNTBull2} and finite order
higher derivative $\alpha^{\prime}$-contributions
\cite{Cardoso,CardosowTalk,0611140} into the black hole entropy. It is expected that
there would exist a generic $\alpha^{\prime}$-corrected
entropy formula, which would offer an intrinsic account of the
statistical models arising from associated microscopic conformal
field theories. At this point, it is worth mentioning that the
interesting attractor inter-relations and their microscopic
details are beyond the scope present set-up. Thus, we leave these
issues for a future exploration.

Finally, it is worth mentioning that the higher fluxes are
expected to play an important role in understanding the physics of
black holes. In such cases, the moduli space geometry, Sen entropy
function, flow equations and the factorization properties, e.g.
Eqn.$\ref{moduli}$, may be explored further at or way from the
attractor fixed points. Further introspecting seems interesting
towards an unified understanding of the entropy functions with an
inclusion of non-zero higher fluxes.  
\section*{Acknowledgement}
BNT would like to thank Prof. V. Ravishankar and Prof. S. Bellucci
for support and encouragements; and Dr. Md. A. Bhat, Dr. V.
Chandra for comments and suggestions, while this work was under
preparation. BNT would further like to thank the organizers of
\textit{``National Conference on New Trends in Field Theories,
(1st-2nd Nov. 2008), Department of Physics, Banaras Hindu
University, Varanasi, India''} for providing stimulating
environment towards this work. The research of BNT has been
supported in part by (i) \textit{``CSIR, New Delhi, India''}
(under the fellowship \textit{``CSIR-SRF-9/92(343)/2004-EMR-I''}),
(ii) \textit{``Indian Institute of Technology Kanpur, India''} and
(iii) \textit{``INFN-Laboratori Nazionali di Frascati, Roma,
Italy''}.

\end{document}